\newcommand{\beq}{\begin{equation}}
\newcommand{\eeq}{\end{equation}}
\newcommand{\bea}{\begin{eqnarray}}
\newcommand{\eea}{\end{eqnarray}}
\newcommand{\rd}{\rho_{\rm d}}
\newcommand{\bd}{\beta_{\rm d}}
\newcommand{\tmax}{T_{\rm max}}
\newcommand{\tm}{t_{\rm max}}

\documentstyle[aps,prb,multicol,graphicx]{revtex}
\begin{document}
\draft

\title{Metal-insulator transition in 2D: resistance
 in the critical region}

\author{B.\ L.\ Altshuler$^{1,2}$, D.\ L.\ Maslov,$^{3}$
 and V.\ M.\ Pudalov$^{4}$}
\address{
$^{1)}$NEC Research Institute, 4 Independence Way, Princeton, NJ
08540\\ $^{2)}$Physics Department, Princeton University,
Princeton, NJ 08544\\ $^{3)}$Department of Physics, University of
Florida\\ P.\ O.\ Box 118440, Gainesville, Florida 32611-8440\\
$^{4)}$P.\ N.\ Lebedev Physics Institute, Russian Academy of
Sciences\\ Leninsky Prospect 53, Moscow 117924, Russia }
\maketitle \centerline{(\today)}
\vspace{0.3in}
\begin{abstract}
The goal of this paper is to highlight several issues which are
most crucial for the understanding of the \lq\lq metal-insulator
transition\rq\rq\/ in two dimensions. We discuss some common problems in
interpreting experimental results on high mobility Si MOSFETs. We
analyze concepts and methods used to determine the critical
density of electrons at the metal-insulator transition. In
particular, we discuss the origin of the temperature dependence of
the resistivity and reasons for this dependence to flatten out at
some electron density in the vicinity of the metal-insulator
transition. This flattening has recently been proposed to indicate
a  true quantum phase transition. We suggest an alternative
interpretation of this result and demonstrate the consistency of
our proposition with the experimental data.
 One of the main questions, which arise
in connection with the transition, is whether or not the metallic
state is qualitatively distinct from a conventional disordered
Fermi liquid. We analyze the arguments in favor of both
affirmative and negative answers to this question and conclude
that the experimental results accumulated up-to-date do not
provide convincing evidence for the new  state of matter
characterized by a metallic-like residual conductivity. We also
discuss in details the measurement and control of the {\it
electron} temperature; these issues are crucial for interpreting
the low-temperature experimental data.
\vspace{0.3in}
\end{abstract}
\newpage
\begin{multicols}{2}
\section{Introduction}
\label{sec:intro}
After several years of intensive experimental and
theoretical efforts (see, e.g., Ref.~[\onlinecite{ensslin}]
for an extensive bibliography), even the basic features of the
phenomenon known as the  \lq\lq metal-insulator transition in two
dimensions\rq\rq (2D MIT) remain to be the subjects of ongoing
discussions.
Is this phenomenon a true \lq\lq quantum phase transition\rq\rq  (QPT)
[\onlinecite{sondhi}] or
can it be
understood in terms of conventional physics of disordered
conductors [\onlinecite{leeandrama,aa}]~?
This question is at the heart of the whole discussion.
Recently, we wrote a paper
[\onlinecite{amp}] in favor of the second possibility.
In particular, we argued that
\begin{enumerate}
\item[i)]~in
the \lq\lq metallic phase\rq\rq\/, 2D systems seem to behave
as a quite conventional disordered metal rather than a distinctly new
state of matter and
\item[ii)]~ it is possible to explain the anomalous
behavior of  the resistivity, $\rho(T)$, by an interplay of two
temperature dependences: the one given by
a metallic-like, quasiclassical (Drude) resistivity
[\onlinecite{comment}],
$\rd(T)$,
and the other one arising from quantum localization.
\end{enumerate}

We do realize that
ii)~ represents a rather naive approach, at least
because it does not fully take into account
the Coulomb interaction between
electrons, whereas the most pronounced effects have been observed
in systems with {\it a priori} strong electron-electron correlations.
Nevertheless, this approach allows one to describe qualitatively, and
even semi-quantitatively,
most of the results on electron transport
both in metallic phase and near the transition
point.
This suggests that it is
a good starting point for developing an adequate theoretical
understanding
 of the MIT
in two dimensions.

Conclusion i) is based on the analysis of
existing experimental data, as
summarized in
Refs.~[\onlinecite{amp,meir1}].
 It has been further supported by recent
experimental papers [\onlinecite{ensslin,gmax,pudalovHall,simmons}],
which showed that the transport properties of Si/Ge, Si MOSFET and p-GaAs
structures in the metallic regime
can be successfully interpreted
in conventional terms. Authors of
Refs.~[\onlinecite{pudalovHall,simmons}] have also identified the
contribution of
electron-electron interactions to the resistance (via
measuring the temperature-dependent part of the Hall resistance) and
found that it remains  quite small even when parameter $r_s$ is rather
large.
There is also a number of  recent theoretical papers
[\onlinecite{dasklap,das98,meir,yaish,meir1}]
which, although differing
in a particular mechanism for the
$T$-dependence of the Drude resistivity,
share the general
spirit of  propositions i) and ii).

An alternative point of view, expressed, for example,
in Refs.~[\onlinecite{kk,sachdev,krav9812389}], is that the
observed MIT-like behavior indicates a true quantum phase transition
between
an insulator and a novel metallic phase.  Motivated by the importance of
this
controversy, i.e., QPT versus propositions i) and ii), for the
field of quantum transport in 2D and its possible
relevance for other realizations of quantum phase transitions, we
decided to analyze in detail the arguments on both sides of the QPT
question.
 In this paper, we discuss recent and some of the previously
published experimental results on the resistivity of Si MOSFETs
in the vicinity of the MIT, along with some of the theoretical
interpretations of these results. We conclude that
that there is no convincing experimental evidence for the QPT
in the existing data.

\begin{figure}
\begin{center}
\resizebox{3.2in}{5.in}{\includegraphics{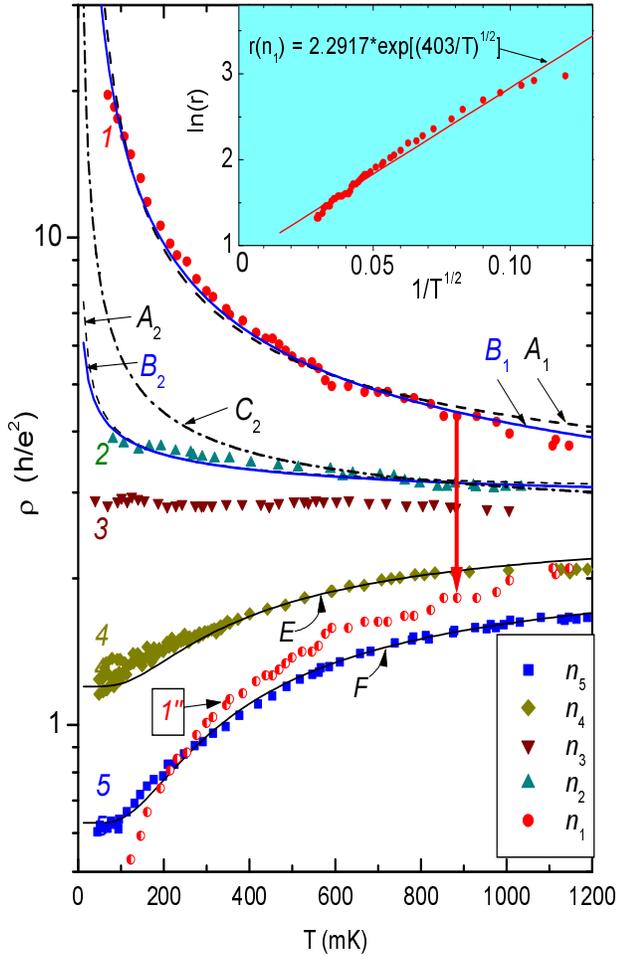}}
\begin{minipage}{3in}
\vspace{0.1in} \caption{Resistivity vs temperature, reproduced
from Fig.~2 of Ref.~[15]. Electron density $n$ in units of
$10^{10}$\,cm$^{-2}$: {\it 1} - 6.85, {\it 2} - 7.17, {\it 3} -
7.25, {\it 4} - 7.57, {\it 5} - 7.85. Dashed curves $ A_1$ and
$A_2$: fits of data {\it 1} and {\it 2} by $2.29
\exp\left[(403/T)^{1/2}\right]$  and $2.84
\exp\left[(11.11/T)^{1/2}\right]$, respectively ($T$ is in mK).
Solid curves $B_1$ and $B_2$: fits of the same data by
$1.25\exp\left[(1685/T)^{1/3}\right]$ and
$2.55\exp\left[(8./T)^{1/3}\right]$, respectively. Semi-open
symbols ($1''$): reflection of data points {\it 1} about curve {\it
3}. Curves $E$ and $F$: best fits of data {\it 4} and {\it 5} by
$\rho_4(T) = 1.2 + 1.45 \exp(-450/T)$ and $\rho_5(T) = 0.63 + 1.6
\exp(-485/T)$, respectively.}
\label{fig1}
\end{minipage}
\end{center}
\end{figure}

As one of the main questions in the field is the behavior of a 2D system
in the limit $T\to 0$,
 we begin our analysis
with a short summary of
recent results by Kravchenko and Klapwijk (KK) [\onlinecite{kk}],
who measured the temperature dependence of the resistivity down
to a
bath temperature as low as $T_{\rm bath} =35$\,mK.
This paper reports measurements
on a single Si MOSFET sample with peak mobility $\mu_{\rm
peak}=27,000$\,cm$^{2}$/Vs
in
the range from 35\,mK to 1.2\,K at five different
electron concentrations.
Digitized data from Ref.~[\onlinecite{kk}]
is shown in Fig.~\ref{fig1} (curves {\it 1-5}).

In Ref.~[\onlinecite{kk}],  the following two points are emphasized:
\begin{enumerate}
\item[1)]
For $n=n_{1}=6.85 \times 10^{10}$cm$^{-2}$ and $n=n_{2}=7.17
\times 10^{10}$cm$^{-2}$ (curves {\it 1} and {\it 2}), the
resistivity decreases as temperature increases $(d
\rho/dT<0)$; for $n=n_{4}=7.57\times 10^{10}$cm$^{-2}$ and
$n=n_{5}=7.85 \times 10^{10}$cm$^{-2}$ (curves {\it 4} and {\it
5}), the resistivity increases with temperature
$(d\rho/dT>0)$. Based on these observations, the
authors argue that curves {\it 1} and {\it 2} correspond to an
insulating phase, whereas curves {\it 4} and {\it 5} demonstrate a
metallic behavior. At $n=n_{3}=7.25 \times 10^{10}$cm$^{-2}$
(curve {\it 3}), only a weak ($\pm 5\%$) temperature dependence
was observed within the interval of bath temperature from 35\,mK
to 1\,K. The authors conclude that  $n_{3}=7.25 \times 10^{10}
$cm$^{-2}$ is the critical electron density for their
sample, i.e., it corresponds to the metal-insulator transition
point. Note that this conclusion is based entirely
on
the temperature independence of curve {\it 3}
over a relatively narrow temperature range.
\item[2)]
Another observation,
emphasized in Ref.~[\onlinecite{kk}], is that
in the range
0.1\,K$<T<$\,0.4\,K the metallic
$\rho(T)$ - dependences (curves {\it 4, 5}) appear to be nearly
linear
 and thus
different from the exponential
temperature dependence observed in other experiments.
KK
stress that in this region
the
resistivity shows no insulating up-turn at the lowest temperature
achieved.
\end{enumerate}

KK
consider the fact that they observed no temperature dependence
of the resistivity of their sample at a certain concentration
$n=n_{3}$ as a strong argument against our
model [\onlinecite{amp}]. They point
out that such a precise cancellation would require a special
relation between $\rho_{d}(T)$ and the scaling $\beta (\rho
)$-function. KK argue that this relation between two objects of
entirely different origin \lq\lq would be a remarkable
coincidence\rq\rq\/, and is therefore unlikely.

Based on the arguments listed above,
authors of Ref. [~\onlinecite{kk}]
conclude that their
experiments are
consistent with the existence of
the zero-temperature quantum phase transition
.

\begin{figure}
\begin{center}
\resizebox{3.2in}{5.1in}{\includegraphics{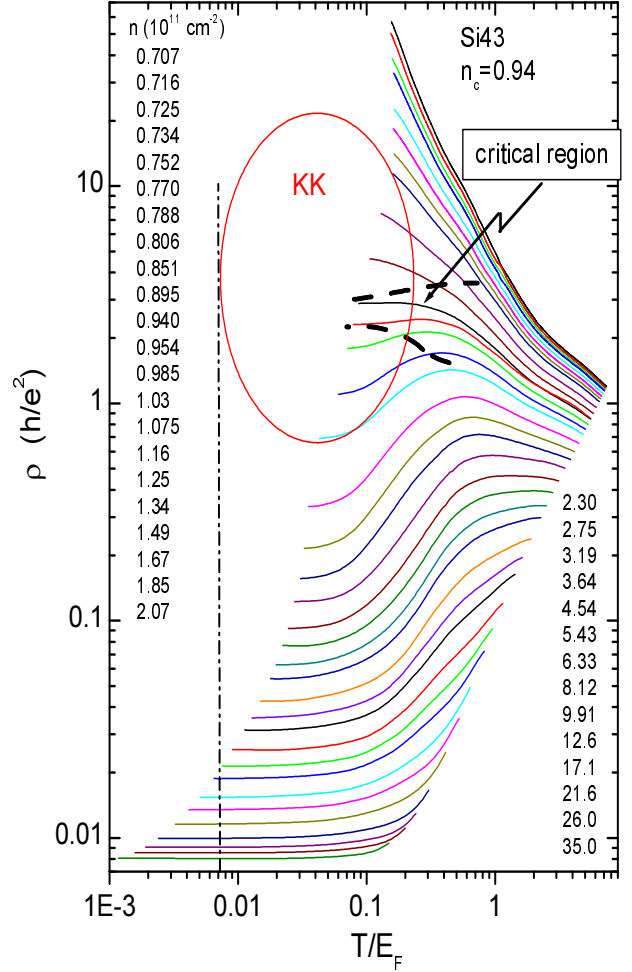}}
\begin{minipage}{3in}
\vspace{0.1in}
\caption{Resistivity vs temperature (from 0.3 to
$\approx 40$\,K) for a wide range of densities and Fermi energies. The
figure is reproduced from Ref.~[18]. Encircled region: domain
of parameters explored in Ref.~[15].
Dash-dotted vertical line depicts the empirical temperature,
$T_Q=0.007E_F$, below which the logarithmic temperature dependence
sets in.}
\label{fig2}
\end{minipage}
\end{center}
\end{figure}

\vspace{-0.1in}
Note that a rather narrow range of densities was explored
in Ref.~[\onlinecite{kk}].
To show
the development of the
MIT-phenomenon
over
a wider range of densities and temperatures,
we present in
Fig.~\ref{fig2} the results of Ref.~[\onlinecite{hawai99}]
for a sample with a close value of $\mu_{\rm peak}$.
The encircled region of the $\{T,n\}$ plane
(region KK in this figure) indicates roughly the domain of parameters explored
 in Ref.~[\onlinecite{kk}].
The temperature in Fig.~2 is normalized to the Fermi energy
$E_F$, in
order to demonstrate that
resistivities corresponding to quite different Fermi energies, and
therefore
$r_s$-values, exhibit similar $T$-dependences. A strong (50-100\%)
drop in $\rho$ is still present for
$n\lesssim 20\times 10^{11}$\,cm$^{-2}$, which corresponds to $r_s\sim 2$ and
$\rho\sim 0.01h/e^2$.
 Thus, contrary
to the popular opinion,
the resistance drop is not intrinsic only to the low density range
(high $r_s$-values).

\vspace{-0.1in}
\vspace{-0.1in}
We would like to stress once again that in the region
of high densities
and small resistivities quantum interference corrections of any kind
(weak localization, exchange and Hartree interactions, spin-orbit, etc.)
would amount only to a few percent variation (of both signs) in $\rho$
with $T$.
The phonon contribution to the resistivity
of Si-MOSFETs
can be shown to be
negligible at least for
$n\lesssim20\times 10^{11}$\,cm$^{-2}$ in the relevant $T$-range. Therefore,
there must be some other mechanism of the $T$-dependence at work, neither
of quantum interference nor of phonon nature.
The resistance
drop of non-phonon nature persists up to
the highest temperature of these measurements,
which is $\approx 35 - 40$\,K
for each curve in Fig.~2
(for conversion of densities into the Fermi-energies, use the formula:
$E_F$[K]$ = 7.31\times n [10^{11}$\,cm$^{-2}]$).

\section{Metal, Insulator, and Critical Point. General Discussion}
\label{sec:mitgen}
\vspace{-0.1in}
Contrary to a common belief, a metallic phase and thus
a metal-insulator transition can in principle occur in two
dimensions. For example, in a hypothetical situation of non-interacting
electrons spin-orbit coupling results in resistivity vanishing with
temperature,
provided that disorder is sufficiently weak (\lq\lq weak
antilocalization\rq\rq\/ [\onlinecite{leeandrama,aa,hikami}]).
However, if disorder is strong enough, it leads to localization.
The metallic state is stable with respect to weak and short-ranged
electron-electron interactions. The Coulomb interaction between electrons
for
small $r_s$ overwhelms weak antilocalization and thus destroys the
metallic state with zero residual resistivity. Nevertheless, the Coulomb
interaction
itself  brings in an antilocalizing contribution to the resistivity
(triplet channel). For small $r_s$, this contribution is
smaller than the one from
the singlet channel, and thus the net effect of
interactions is the
enhancement of localization. It may
be the case that the triplet  channel wins for larger
$r_s$ and a metallic
state becomes possible. (In fact, this scenario follows
from the RG treatment
[\onlinecite{Fin,dicastro,belitz_kirkpatrick}].)
However, if such
a state exists, it
must be distinct from the conventional \lq\lq disordered Fermi
liquid\rq\rq\/.

So far, the experimental evidence is in contrast with the existence
of a distinctly new state
(see Refs.[\onlinecite{amp,meir1}]). In fact,
at higher densities
the resistivity exhibits an insulating up-turn in agreement
with the weak localization theory.
We believe that in the existing experiments the true low-temperature
asymptotic
behavior has not yet been achieved for lower densities (in the vicinity
of the transition). Therefore, one
cannot interpret the crossover between
$d\rho /dT >  0$  and $d\rho /dT <  0$ at finite temperatures
as a proof for the existence of two distinct phases.
Nevertheless, it is instructive to adopt such an
interpretation for a moment and to compare the experimentally observed
$\rho (T)$ with the commonly accepted picture of a quantum phase
transition.
Let us first recall what is known about the transition in 3D.

We start our
analysis
by formulating
the definitions of metal and insulator.
Arguing about definitions is not a too rational thing to do -- a
definition cannot be
right or wrong.
We simply point out that for the well-studied case of MIT in 3D,
the commonly held definitions of both phases differ from
those adopted by KK.
In 3D systems, $\rho_{\rm insulator}(T\to 0) \to\infty$, whereas
$\rho_{\rm metal}(T\to 0)$ is finite (see
[\onlinecite{leeandrama,aa,belitz_kirkpatrick}]
and references therein).
In fact, $d\rho/dT$ is negative  in both the metallic and
insulating phases, provided that the system is close to the transition
point and the temperature is low enough.

For non-interacting particles, the conventional scenario of the 3D
MIT is well supported by the perturbation theory and renormalization
group arguments
[\onlinecite{leeandrama,aa,gang4,belitz_kirkpatrick}].
According to this scenario, in the metallic phase close to the transition
$\rho(T)$ increases as $T^{-\alpha}$ as temperature decreases.
The increase of $\rho(T)$ saturates at $T \lesssim T_{\rm  sat}$. Upon
approaching the critical point, $T_{{\rm sat}}$\ tends to zero.
Therefore, exactly at the critical density the
$\rho(T)$-dependence is a power law:
$\rho(T) \sim T^{-\alpha}$.
In the insulating phase, the resistivity behaves as
$\exp(T_{0}/T)$ (nearest neighbor hopping) or as
$\exp[(T_{0}/T)^{\beta}]$ (variable range hopping).
The exponent $\alpha$ is determined by mechanisms of dephasing. As a
result, it is, strictly speaking, non-universal.
Let the dephasing rate scale with temperature as
$1/\tau_{\varphi } \propto T^p$.
Then assuming that (a) one-parameter scaling holds,
(b) this parameter is the conductance, and (c) $p=1$,
one arrives at $\alpha = 1/3$.
Note that the
electron-electron interactions can change the value of $\alpha$.

Numerous experiments on 3D doped semiconductors confirmed the power-law
behavior of $\rho$ in the critical region. However, there is still no
consensus regarding the value of $\alpha$:
both $\alpha= 1/2 $ [\onlinecite{onehalf1,onehalf2,onehalf3}]
and  $\alpha = 1/3$  [\onlinecite{onethird1,onethird2}] have been
reported. It is also possible  that $\alpha$  depends on whether a
semiconductor is compensated or not.
(For the review of an MIT in 3D see Ref.~[\onlinecite{sarachik}].)

Returning to the MIT in 2D, we note that
the assumption of a
temperature-independent
resistivity at the critical
density is as doubtful as the statement
that $\alpha = 1/3$ in 3D.
Indeed, both of these points can be justified only within the one-
parameter scaling.
This scaling does not seem to apply universally even in the 3D case. For
a 2D system of noninteracting electrons, this scaling predicts no
metallic state at all. To get a chance to describe a metallic state, one
needs to add
more ingredients, e.g., electron-electron
interactions, to the theory.
Once we deal with a problem which is richer than localization in
quenched disorder, there are no reasons to assume that the one-parameter
scaling still holds
(see, e.g., Ref.~[\onlinecite{Fin}]).
Therefore, a temperature-independent critical resistivity is an
assumption rather than a law of nature.

As an example, consider the following density- and
temperature-dependences of the resistivity (which do not follow from any
of the existing theories, but do not contradict to any of the commonly
respected general principles either):
\begin{equation}
\rho(n,T)=\rho_{0}(n)+\rho_{1}(n)\left[\frac{T}{T_{1}}\right]^{-\alpha}
\exp \left[\frac{T_{0}(n)}{T}\right].
\end{equation}
It is natural to
define
the critical concentration, $n_{c}$,
as the concentration at which $T_{0}(n)$ changes sign:
$T_{0}(n_c)=0$; for $n>n_{c}$ (metal), $T_{0}(n)$ is negative,
whereas for $n>n_{c}$  (insulator), $T_{0}(n)$ is positive.
It is also assumed that $\rho_{0}(n)$,\ $\rho_{1}(n)$ and $T_{1}(n)$
are smooth functions of the density in the vicinity of the transition
point $n=n_{c}$.
According to Eq.(1), exactly at the critical point ($n=n_{c}$)
\begin{equation}
\rho(n_{c},T)=\rho_{0} + \rho_{1}\left(\frac{T}{T_{1}}\right)^{-\alpha}.
\end{equation}

As we see, the critical resistivity is temperature-dependent and can even
diverge as $T \to 0$.
At the same time, the resistivity saturates at $\rho=\rho_{0}(n)$
for $T < -T_{0}$ in the metallic regime.
(In our example, $\rho (T)$
has a maximum at $T= -T_{0}/\alpha$.
We do not think that such a maximum
is a mandatory feature
of the MIT in two
dimensions.)
We conclude that, at least in this example, a temperature-independent
resistivity is a signature of a metal rather than of a critical point.

Regardless of this particular and rather artificial example, we notice
that it is neither easy nor a straightforward task to determine the
critical point of an MIT.
Quantum phase transitions are zero temperature phenomena, whereas
experiments are performed at finite $T$. Therefore, to determine the critical
concentration, one does not have another choice but to analyze the data
taken at the lowest accessible temperatures.
This analysis should prove that $\rho(T)$ indeed acquires
{\it insulating exponential behavior} as soon as the concentration gets
lower than the apparent value of $n_{c}$.

We find it more appropriate to use the onset of the insulating
exponential behavior in $\rho (T)$ rather than the vanishing of  the
derivative $d\rho/dT$ for an experimental
identification of the critical point.
We are not aware of any reasons to assign the meaning of the critical
point to a density $n^*_c$ at which the temperature dependence of the
resistivity is least pronounced, as it was done by KK.
Indeed, direct measurements  of the two quantities, $n^*_c$ (from the
temperature independent  \lq\lq separatrix\rq\rq\/) and $n_c$ (from the
disappearance of the non-Ohmic hopping behavior) confirm their systematic
difference for high-$\mu$ samples:
$n^*_c$ is larger than $n_c$ by 1-5\% (see, e.g., Fig.~3 in
Ref.~[\onlinecite{noscaling}]).
A similar difference
arises also from the comparison of
critical behaviors in the thermoelectric power and conductance
[\onlinecite{TEPMIT}]. Note that four out of five electron densities
in Ref.~[\onlinecite{kk}], $n=n_1-n_4$, fall into this ambiguous
interval.\\
\vspace{-0.3in}
\section{Experimental determination of the critical point.}
\vspace{-0.1in}
We now turn to the experimental data of Ref.~[\onlinecite{kk}].
The authors assume that $n_{c}$ equals to $7.25 \times
10^{10}$cm$^{-2}$
(curve {\it 3} in Fig.~\ref{fig1}). According to
the definition of the critical point, proposed in Sec.~\ref{sec:mitgen},
this assumption implies that at lower densities ($n=n_{2}=7.17 \times
10^{10}$cm$^{-2}$ and $n=n_{1}=6.85 \times 10^{10}$cm$^{-2}$), the
resistivity increases exponentially as temperature decreases.
It turns out that one can fit neither $\rho(T,n=n_{2})$ nor
$\rho(T,n=n_{1})$
with a simple exponential dependence.
However, $\rho(T,n=n_{1})$ can be satisfactorily  approximated by
\begin {equation}
\rho_{\rm exp}(T,n) = \rho^*_m
\exp \left[\left(T_{0m}(n)/T\right)^m\right]
\label{expfit}\end {equation}
with either $m=1/2$ or $m=1/3$ (dashed line $A_1$ and solid line $B_1$ in
Fig.~\ref{fig1}, respectively). The quality of the fit can be seen from
the deviations of the experimental data from the straight line in the
inset to Fig.~\ref{fig1}.
Although curve {\it 2} in Fiq.~\ref{fig1} does not seem to behave
exponentially, one can still try to fit it by function
(\ref{expfit}) in two different ways:
\begin{enumerate}
\item[i)]
one can assume that $T_{0m}$
is proportional to $|n-n_c|$. (This dependence was observed
in previous studies of the MIT in Si
MOSFETs [\onlinecite{krav94,krav95,krav96}]).
Then, using the value of $T_{0m}$ for $n=n_1$, one determines $T_{0p}$
for $n=n_2$ by linear extrapolation, whereas $\rho^*_m$ is found by
fitting the experimental data. The results of this procedure
for $m=1/2$ are shown in Fig.~\ref{fig1} (dashed-and-dotted line
C$_2$). It is quite clear that such a fit does not work (the same is true
for $m=1/3$, not shown).
\vspace{-0.1in}
\item[ii)] Alternatively, one can simply make the best fit of curve {\it
2} by function (\ref{expfit}) for $m=1/2,1/3$, treating $T_{0m}$ and
$\rho^*_m$ as independent parameters.
Results of this fit
are depicted in Fig.~\ref{fig1} by curves $A_2$
and $B_2$, respectively.
It is customary to assume that in the vicinity of the critical point,
$T_{0m}$ scales as $|n-n_c|^{z\nu}$.
Given the values of $T_{0m}$ for $n=n_{1,2}$,
one can follow the assumption of KK that $n_c=n_3$ and find
\begin{eqnarray}
z\nu=\frac{\left[\ln T_{0m}(n_1)/T_{0m}(n_2)\right]}{\ln
\left[(n_3-n_1)/(n_3-n_2)\right]}=\\
=\left\{
\begin{array}{cl}
2.23\quad{\rm for}\quad m=1/2;\nonumber\\
3.57\quad {\rm for}\quad m=1/3.\nonumber\\
\end{array}
\right.
\end{eqnarray}
 It is quite alarming that these values of the critical exponent are
substantially different from $z\nu=1.2-1.6$ found
in previous studies
[\onlinecite{krav95,krav96}].
\end{enumerate}

It is also often assumed that in the vicinity of the critical point
$\rho(T,n)$ should be  symmetric with respect to reflection about the
separatrix
[\onlinecite{krav96,shahar,simonian97}].
To check if this relation works, we reflect curve {\it 1} ($|\delta
n|=n_3-n_2=0.4\times 10^{10}$\,cm$^{-2}$) about
the would-be separatrix (curve {\it 3}). The reflected curve is
shown by
semi-open circles. Had reflection symmetry worked, the reflected
curve must have been located in between
curves {\it 4} ($\delta n=0.32$\,cm$^{-2}$) and {\it 5}
($\delta n=0.6$\,cm$^{-2}$), closer to curve {\it 4}.
In fact, the reflected curve {\it crosses} both curves
{\it 4} and {\it 5}. Thus by this criterion $n_3$ is not
a critical density.

Unfortunately, for $n=n_{2}$\,  KK present the data
only down to $80$\,mK, whereas for the rest of the densities, both smaller
$(n_{1})$ and larger $(n_{3},n_{4},n_{5})$ than $n_{2}$,
is
shown
down to 35\,mK.
If observed, a substantial increase in $\rho(T,n_{2})$ with cooling
from $80$\,mK to 35\,mK
would provide KK with a strong argument in favor of $n_{3}$
being close to the critical density.

Within the assumption that the data presented in Fig.~\ref{fig1} can
at all be interpreted within a paradigm of a
quantum phase transition,
the contradictions demonstrated above suggest at least one of the
following three conclusions:
\begin{enumerate}
\item[i)]~the separatrix corresponds not to density
$n_3$ but to some smaller density;
\vspace{-0.1in}
\item[ii)]~the critical exponent is not
universal for a given material but depends on the sample preparation,
geometry, etc.;
\vspace{-0.1in}
\item[iii)]~
the exponential
increase of the resistivity at density $n_2$ is suppressed due
to overheating
(we will discuss this option in Sec.~\ref{sec:heating} in more details).
\end{enumerate}
{\it Alternatively, one may conclude that the data of
Ref. [~\onlinecite{kk}] do not provide evidence
for a quantum phase transition between two distinct states of
matter, but rather indicate a crossover between two different types of
(relatively) high temperature behavior.}
\vspace{-0.2in}
\section{Resistivity of Si MOSFETs in the critical region}
\label{sec:crit}
\vspace{-0.2in}
As it has already been mentioned,
any scenario of a quantum phase transition implies that
for the densities close to the critical point $n_{c}$ the resistivity
$\rho (T,n)$ at sufficiently high temperatures demonstrates a critical
behavior.
This means that the deviation of $\rho (T,n)$ from the separatrix $\rho
(T,n_{c})$ is small for small values of the critical parameter
$|n-n_c|^{z\nu}/T$.
Keeping this in mind, we now recall the main results of earlier
experiments on the resistivity of high mobility Si MOSFETs in the
critical region and compare them with KK's results.

\begin{figure}
\begin{center}
\includegraphics[angle=0,width=3.0in]{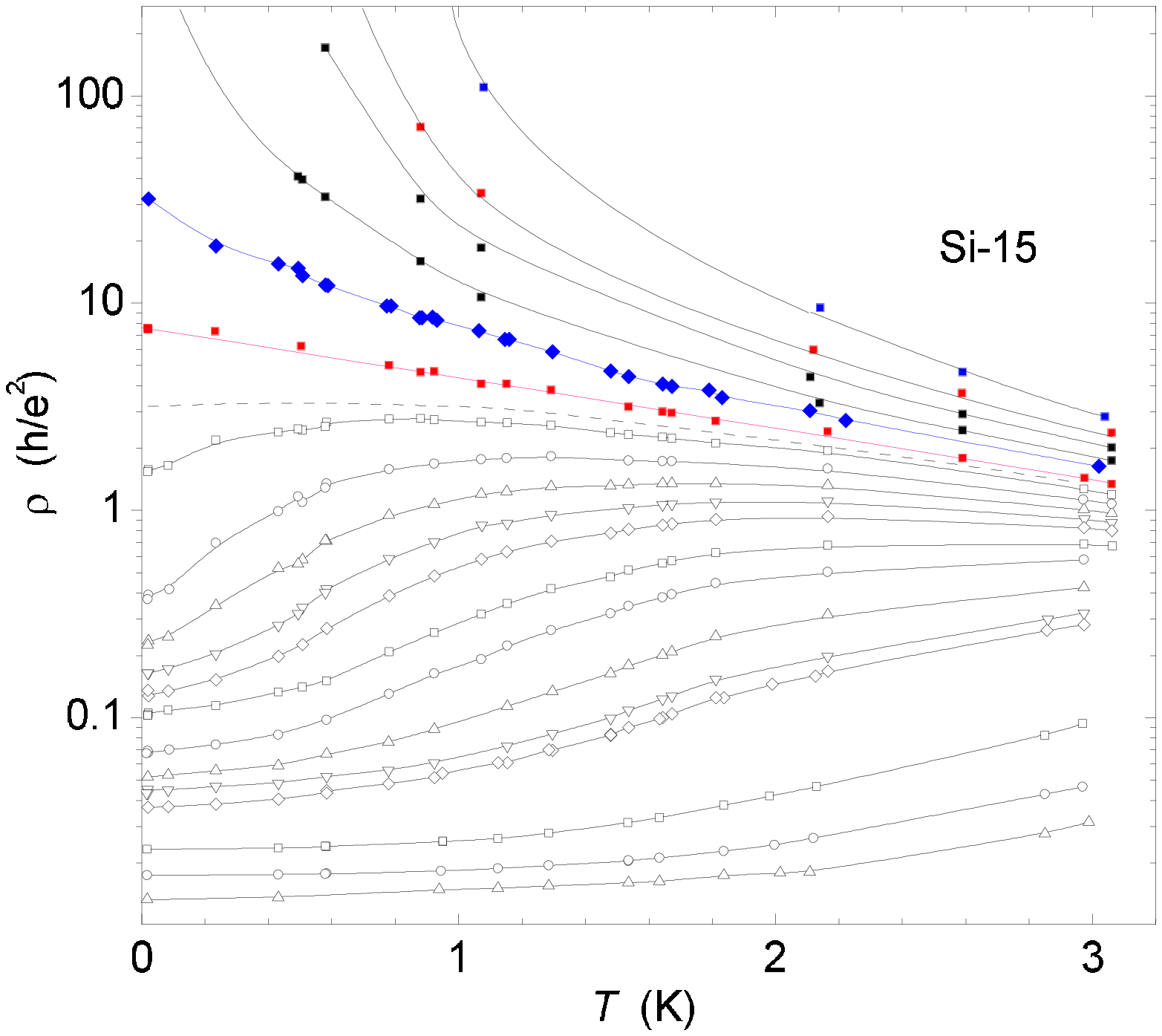}
\begin{minipage}{3.0in}
\caption{$\rho(T)$ over the range from 0.016 to 3K.
Different symbols correspond to $n$ from 0.449 to 0.989 (in steps
of 0.054), and further 1.1, 1.2, 1.42, 1.64, 1.74, 2.82, 3.9,
$4.98\times 10^{11}$\,cm$^{-2}$. Reproduced from Refs.~[30,37].}
\label{fig3}
\end{minipage}
\end{center}
\end{figure}

Consider, for example, the resistivity of a sample
with $\mu_{\rm peak}=55,000$\,cm$^2$/V\,s, reproduced from
Ref.~[\onlinecite{noscaling,mauterndorf}] in Fig.~\ref{fig3}.
Assume for a moment that metallic and insulating behaviors can indeed be
distinguished by the sign of
the derivative
$d\rho/dT$, as suggested by KK.
At first glance, the application of this criterion to the region $T<1$\,K in
Fig.~\ref{fig3}
is rather
unambiguous.
Indeed, $n^*_c$ determined from the condition of $T$-independent  $\rho$
 corresponds to the dashed separatrix.
\vspace{-0.2in}
On the other hand,
using the onset of
activated and non-Ohmic
conduction as a criterion for the transition, we find that the critical density $n_c$
is lower, $n \approx 0.719\times 10^{11}$\,cm$^{-2}$ and
corresponds to
the \lq\lq tilted\rq\rq\/ (6th) curve from the top.
At higher temperatures,
some of the  \lq\lq metallic\rq\rq\/
curves change their slopes and
follow quite closely those \lq\lq insulating\rq\rq\/ curves, which
correspond to lower densities  (see solid curves in Fig.~\ref{fig3}.)
The region of the $\{T,n\}$ plane, where such a behavior occurs, is
defined as the
critical region.
Certainly, the density range of the critical region is narrower for lower
temperatures.

As a rule,
{\it the resistivity depends on the temperature in this region}.
This feature, which can be found in almost any published data,
is demonstrated in Figs.~\ref{fig2} and \ref{fig3}.
For example, $\rho (T,n)$
 in Fig.~\ref{fig3} increases by approximately a factor of 2, as $T$
changes from 3\,K to 1\,K,
both for $n=0.719\times 10^{11}$\,cm$^{-2}$
(the 6th curve from the top)
and  $n=0.773\times 10^{11}$\,cm$^{-2}$
(the 7th curve).

\begin{figure}
\begin{center}
\includegraphics[width=3.in]{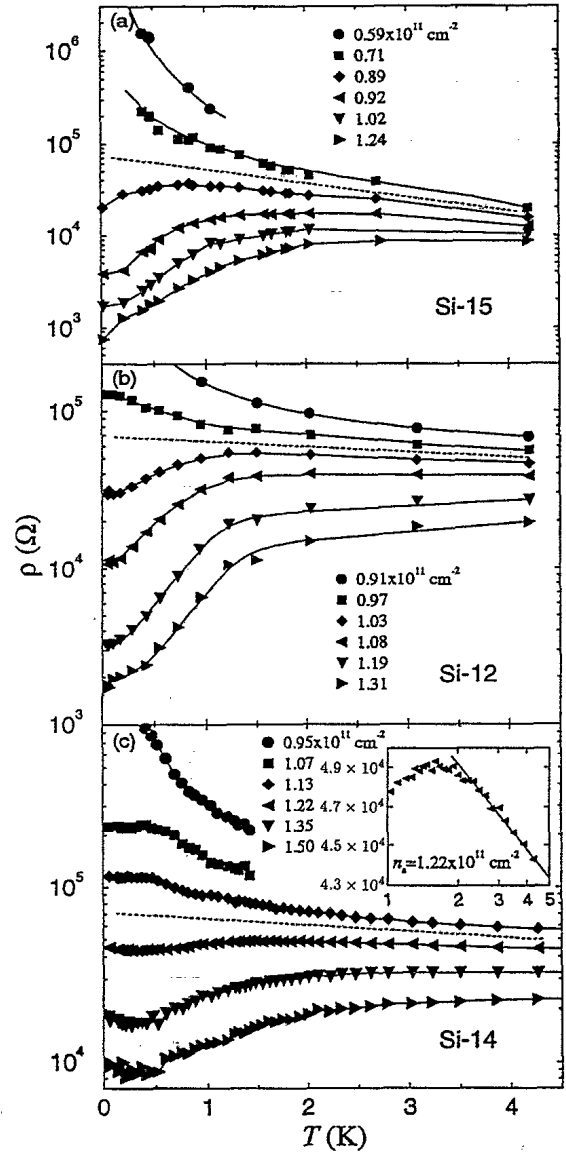}
\begin{minipage}{3in}
\vspace{0.2in} \caption{Resistivity vs $T$ near $n_c$ for 3
samples with mobility decreasing from top to bottom.
Reproduced from Ref.~[32].}
\label{fig4}
\end{minipage}
\end{center}
\end{figure}

\begin{figure}
\begin{center}
\includegraphics[angle=0,width=3.in]{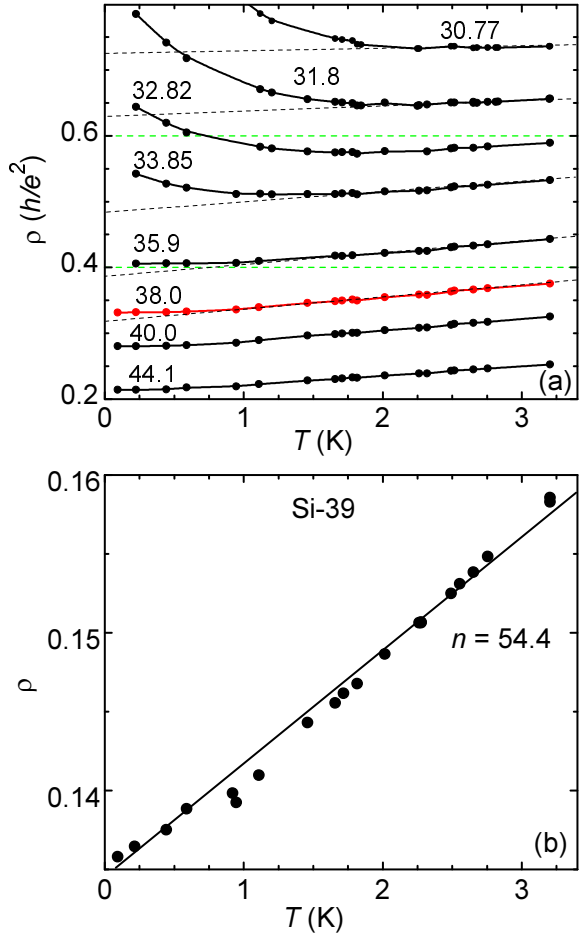}
\begin{minipage}{3in}
\vspace{0.2in} \caption{Resistivity vs temperature for a {\em low
mobility sample} (Si-39). Numbers at the curve indicate the
densities in $10^{10}$cm$^{-2}$. Reproduced from Ref.~[37].}
\label{fig5}
\end{minipage}
\end{center}
\end{figure}

In fact,
the $\rho(T)$-dependence at higher temperatures is usually
more pronounced in higher-mobility samples.
This is illustrated by Fig.~\ref{fig4}, which is
reproduced from
Ref. [\onlinecite {krav94}].
\vspace{-0.1in}
\vspace{-0.2in}
The slope of the dashed line for the sample with $\mu_{\rm peak}=
71,000$\,cm$^2$/Vs (Fig.~\ref{fig4}a) is negative and its absolute
value is about 10 times bigger than
in Fig.~\ref{fig4}b ($\mu_{\rm peak}= 33,000$\,cm$^2$/Vs).
The reduction of the slope with $\mu_{\rm peak}$ persists down to
relatively low mobilities until eventually the slope changes sign.
This is seen from Fig.~\ref{fig5}a, reproduced from
Ref.~[\onlinecite{mauterndorf}], where the authors presented
$\rho (T)$ for a sample with $\mu_{\rm peak}$ as small as $5,000$
cm$^2$/Vs.

Even for this low-mobility sample, the $\rho (T)$-dependence
at
the highest density measured ($n = 54.4\times 10^{11}$cm$^{-2}$) shows no
signs of saturation in the accessible temperature region
(above 100\,mK), see Fig.~\ref{fig5}b.
There is no consistent theoretical understanding of this $\mu_{\rm
peak}$-dependence of the slope yet.
Such an understanding will probably come together with the description
of the \lq\lq critical resistance\rq\rq\/, which apparently changes from
about $0.3 e^{2}/h$ at low mobilities to several units of
$e^{2}/h$ at high $\mu_{\rm peak}$[\onlinecite{noscaling,mauterndorf}],
(see also discussion in Refs.~[\onlinecite{kravcomment,amreply}]).
On the other hand, it is natural that {\it for the intermediate
mobilities, the slope is small}, because it should vanish somewhere in the
range from 5,000 to 30,000\,cm$^{2}$/Vs. The mobility of KK's sample is
27,000\,cm$^{2}$/Vs, i.e., right in this range. Therefore one should not
be too surprised  by the fact that {\it in some range of the
concentrations the temperature dependence of the resistivity is rather
flat}, as it was observed by KK.

 Last but not least: a variety of the $\rho(T)$- dependences
in the critical region indicates by itself the absence of universality in
observed
MITs [\onlinecite{noscaling}]
and raises doubts that Si MOSFETs undergo a genuine quantum phase
transition.

\section{\lq\lq How Cold are the Electrons?\rq\rq\/}
\label{sec:heating}
Heating of electrons by the applied source-drain field
as well as  by external noise is a common
problem in low-temperature transport measurements.
This problem turns out to be especially serious
in Si MOSFETS in the vicinity of the
\lq\lq metal-insulator transition\rq\rq\/  and at temperatures
$\lesssim  100$\,mK due to *) weak
temperature dependence of the resistivity in this range of densities and
**) weak electron-phonon coupling.
As a result, the interpretation of resistivity measurements
becomes rather ambiguous.
For example, the temperature
interval, in which the resistivity appears to be temperature-independent,
may look to be much broader than it actually is. The goal
of this section is to demonstrate the seriousness of this problem.

In macroscopic Si MOSFETs, as well as in bulk Si, electrons  couple to
phonons only via the deformation potential.
As this coupling is rather weak, the electron temperature, $T_e$, may
exceed  substantially the bath temperature, $T_{\rm bath}$;
it is also not easy to
control $T_e$.
To provide convincing data, one has to
use a \lq\lq thermometer\rq\rq
capable of measuring $T_e$ directly.
Any observable which depends on $T_e$, other than the
(zero-magnetic-field) resistivity itself,
can be used for this purpose.
In earlier studies on high mobility Si-MOSFETs,
Refs.~[\onlinecite{noscaling,mauterndorf,prl93,pudprb92,JETPLlnT}],
which reported results of the resistivity measurements down to
$T_{\rm bath}=18-25$\,mK,
the electron temperature
was attempted to be measured via\\
i)~the amplitude of Shubnikov - de Haas (ShdH)
oscillations,\\
ii)~temperature dependence of the hopping resistivity,\\
iii)~linearity of $I-V$-characteristics,
and \\
iv)~$T_e$-dependence of the phase relaxation
time.\\
Unfortunately,
results on $T_e$ in the up-to-date
measurements of 2D MIT
are
neither fully consistent
with each other
nor convincing
enough
below $T_{\rm bath} =250$\,mK.

No independent measurement of $T_e$ is reported in Ref. [\onlinecite{kk}].
It is only mentioned
that the source-to-drain bias, $U$,  was less than
200\,$\mu$V
\lq\lq to
ensure that the total power dissipated in the  sample was less
than $10^{-13}$\,W\rq\rq\/.
As it is not clear
whether
this is enough to prevent
electrons from being overheated, it is worth discussing
electron heating in Si MOSFETs.

Electrons are driven out of the equilibrium by the applied voltage and/or
by external noise.
Let $P$ be the power deposited in the electron system.
In the stationary state, all this power leaves the system either with
electrons (through contacts) or with phonons.
The phonon mechanism dominates at higher temperatures.
If $T$ is low enough, this mechanism can be neglected compared to
electron out-diffusion, in which
thermal conductivity of the electrons determines the
heat balance.
As our task is to estimate heating at $T \sim $100\,mK, we first
discuss what happens  without phonons.

Given the temperature increment $\Delta T= T_e - T_{\rm bath}$,
the power, which
is carried out by electrons
through the leads, can be estimated as
(see Refs.~[\onlinecite{prober}] and also
 [\onlinecite{aag}] for
recent discussion):
\begin{equation}
P = \left(\frac{2\pi }{e}\right)^2 \frac{T_{\rm bath} \Delta T }{R},
\label{power}\end{equation}
where $R$ is the resistance which differs from the resistivity by the
aspect ratio.
To obtain a lower bound estimate of the bath temperature
$T_{\rm h}$ at which heating becomes strong, i.e., $\Delta
T\simeq T_{\rm h}$, we
neglect heating due to external noise and
assume that the main reason for heating is the source-to-drain bias,
$U$.
Accordingly,  $P = U^2/R$.
Expressing the ratio $\Delta T/T_{\rm bath}$
through
$P$ via Eq.~(\ref{power}), we obtain
\begin{equation}
\frac{\Delta T }{T_{\rm bath}} = PR \left(\frac{e}{2\pi T_{\rm
bath}}\right)^2 =
\left(\frac{eU}{2\pi T_{\rm bath}}\right)^2.
\label{power1}\end{equation}

Strictly speaking, Eqs.(\ref{power},\ref{power1}) are valid only when
$\Delta T \ll T_{\rm bath}$.
Nevertheless, one can use them to estimate $T_{\rm h}$ as
\beq
T_{\rm h} [{\rm mK}]= 1.8U[\mu{\rm V}].
\label{th}\eeq
For $U\sim 200\mu $V, Eq.~(\ref{th}) gives $T_{\rm
h}=360$\,mK.
For lower bias, e.g., for $U=50\,\mu$V this estimate gives
$T_{\rm{h}}=90$\,mK, whereas the experimental dependence $T_{e}(P)$
[\onlinecite{prober_prb}] taken at
$T_{\rm bath}=140$\,mK shows that at this bias electrons are already
overheated: $T_e=1.4T_{\rm bath}$.
Despite its simple form, Eq.~(\ref{th}) is quite general
and universal. For example, it is also valid
in the presence of Fermi-liquid interactions, which
do not affect the Wiedemann-Franz law [~\onlinecite{reizer}].

Taking into account the phonon mechanism of energy relaxation does not
change our estimate of characteristic temperature $T_{\rm h}$
substantially.
This is so because $T_{\rm h}$
is low enough for out-diffusion
to dominate over phonon emission.
Indeed, the electron-phonon energy loss rate for weak overheating can be
written as
 \beq P_{\rm ph}=\frac{\pi^2}{3}\nu_F LW\frac{T_{\rm
bath}\Delta T}{\tau_{\rm e-ph}}, \label{peph} \eeq where $\nu_F$
is the density of states at the Fermi level, $L$ and $W$ are
the sample length and width,
respectively, and $\tau_{\rm e-ph}$
is the electron-phonon relaxation time.
We assume that electrons are coupled to phonons via the screened
deformation potential.
The corresponding relaxation time for $T_{\rm bath}\ll\hbar k_F
s=3.65\sqrt{n}$\ [K]  (where $s$ is the speed of sound, and $n$ in
$10^{11}${\rm cm}$^{-2}$) can be written as [\onlinecite{levinson,karpus}]
\beq \frac{1}{\tau_{\rm e-ph}}=
\frac{120}{\pi}\frac{\Xi^2m^*k_F^2a_0^3} {\hbar^2Ms}\left(\frac{T_{\rm
bath}}
{k_Fs}\right)^3\left(\frac{T_{\rm bath}}{\kappa s}\right)^2,
\eeq
where $\Xi$ is the deformation
potential constant, $M$ is the atomic mass, $a_0$ is the effective
lattice constant defined in such a way that $M/a_0^3$ is equal to the
mass density, and $\kappa$ is the inverse screening length.
Comparing (\ref{power}) to (\ref{peph}), one finds that
out-diffusion dominates, provided that $T_{\rm bath}$ is smaller
than
\beq T^*=\frac{\pi}{5}\hbar k_Fs\frac{h}{\rho e^2}{\cal E}{\cal P}
r_s^2\frac{\Xi}{\varepsilon _F}\left(\frac{a_0}{L}\right)^2,
\label{tstar0}
\eeq
where ${\cal E}\equiv Ms^2\left(\hbar^2/m^*a_0^2\right)^2/\Xi^3$ and
${\cal P}\equiv\hbar/m^*sa_0$.
Substituting material constants for a Si MOSFET  into (\ref{tstar0}), one
obtains
\beq
T^*[{\rm K}]=0.3\left(\rho n^2\right)^{-1/5} L^{-2/5},
\label{tstar}
\eeq
where $\rho$, $n$, and $L$ are measured in $h/e^2$, $10^{11}$\,cm$^{-2}$,
and mm, respectively.
Using experimental results on electron heating in Si MOSFETs
[\onlinecite {prober_prb}], one estimates the observed
crossover temperature as 0.7 K,
whereas Eq.~(\ref{tstar}) gives $T^*=0.6$\,K
for the experimental conditions of Ref.~[\onlinecite {prober_prb}].
For the conditions of another experiment on heating [\onlinecite{pudphon}],
Eq.~(\ref{tstar}) gives  $T^*=0.3$\,K,
which is just the lowest temperature of this measurement.
No clear crossover from the phonon to out-diffusion mechanisms
was observed in Ref.~[\onlinecite{pudphon}] for $T\geq 0.3$\,K
(although the $T$-dependence of the energy-loss rate does start
to slow down at $T\simeq 1$\,K). This is again consistent with our
estimate. We thus conclude that Eq.~(\ref{tstar})
is consistent with the experimental data
and can serve as at
least a lower bound for $T^*$.

For typical values of $\rho\simeq e^2/h$, $n\simeq 10^{11}$\,cm$^{-2}$,
and $L=0.1-1$\,mm, Eq.~(\ref{tstar}) gives $T^*\simeq 0.3-0.75$\,K, which
is of the same order as $T_{\rm h}$ estimated above as $0.36$\,K.
Note that in deriving (\ref{tstar}) we assumed that $\rho$ was
$T$-independent. Taking the metallic-like $\rho(T)$-dependence
into account enhances out-diffusion at low temperatures
and thus shifts $T^*$ towards even higher values.
Also, taking into account possible external noise would further
increase the value of $T_{\rm h}$.
The lowest bath temperature in Ref.~[\onlinecite{kk}] is $T_{\rm
b}=35$\,mK.
Using Eq.~(\ref{th}), we see that in order to prevent
electron heating, the bias voltage has to be much smaller
than 20 $\mu$V, which is ten times smaller than
the upper boundary for $U$ indicated in  Ref.~[\onlinecite{kk}].

We thus see that there are serious reasons to doubt that the
electron temperature in KK's measurements was
below $\sim 100$\,mK.

\section{Temperature-Independent Resistivity}
\label{sec:tir}
Now it is time to analyze the first of the two main arguments that KK
brought in favor of the \lq\lq true quantum phase transition\rq\rq,
namely, the temperature-independent resistivity at what they
believe to be the critical point.
More specifically, KK claim  that they observed no $T$-dependence (within
a 10\% margin) in the interval $35\,{\rm mK} < T < 1$\,K.
They analyzed the data in terms of our recent theory of Anderson
localization by temperature-dependent disorder [\onlinecite{amp}] and
came to the conclusion that within this theory a nearly constant
$\rho(T)$ would imply a \lq\lq
remarkable coincidence\rq\rq\/.

We begin our discussion of this issue with summarizing briefly
the argument of KK. Afterwards, we present our theoretical
counterarguments.

The theory of Ref.~[\onlinecite{amp}] describes the $T$-dependence of the
observable resistivity, $\rho$, in the presence of two factors: i)~the
$T$-dependence of the Drude resistivity,
$\rho_{\rm d}$, and ii)~Anderson localization. Each of these two
factors brings a $T$-dependence of its own and the resultant
$\rho(T)$-dependence is described by the following scaling equation
\begin{equation}
\left[ \frac{1}{\beta (\rho )}-\gamma\right] \frac{d\ln \rho }{d\ln T}=%
\frac{p}{2}+\left[ \frac{1}{\beta (\rho _{{\rm d}})}-1\right] \frac{d\ln
\rd}{d\ln T},  \label{rgrho}
\end{equation}
where $\rho$ is measured in units of $h/e^2$, $\beta(x)$ is the
Gell-Mann--Low function, whereas $p$ and
$\gamma$ parameterize the phase-breaking time as
\beq
\tau_{\varphi }\propto T^{-p}\rho^{1-2\gamma }.
\eeq
The $T$-dependence of $\rho_{\rm d}$, entering the RHS of
Eq.~(\ref{rgrho}), leads to a variety of behaviors in $\rho(T)$.
In Ref.~[\onlinecite{amp}] we discussed in particular the situation when
$\rho(T)$ exhibits a maximum at  a rather high temperature $T_{\rm max}$
(close to the starting temperature of the RG flow).

Is a very slow variation of $\rho(T)$ over a wide $T$-interval permitted
in this model? KK answer this question negatively.
Their argument goes as follows: to have $d\rho/dT=0$ within some
interval of temperature, one has to require that the RHS of Eq.~(\ref{rgrho})
is equal to zero within this interval. This is possible only if  the
Drude conductivity $\rho_{{\rm d}}$ is related in a specific way both to
the $\beta$-function and to $\tau_{\varphi}$. Such an exact relation is
unlikely, given the different origin of these quantities.

This argument sounds to be formally correct. Nevertheless,
as we will demonstrate shortly,
it is possible to achieve a very small variation of $\rho(T)$
within a wide temperature interval
without imposing any constraints of this kind on $\rd$.
It turns out that the variation of $\rho(T)$ is small enough and the $T$-interval
is wide enough to agree with the experiment.
This possibility results from the fact that there is a whole family of
the $\rho_{{\rm d}}(T)$-functions, which are parameterized
by the electron density and some other parameters, such as
 the peak mobility. It is the freedom in fine-tuning these parameters
that allows one to suppress the variation of the
observable resistivity dramatically.

For the purposes of this section, we adopt the following working
definitions,
consistent with those used in Ref. [\onlinecite{kk}]:
by \lq\lq metal\rq\rq\/ (\lq\lq insulator\rq\rq\/) we
understand the region in which $d\rho/dT>0(<0)$ or $d\rd/dT>0$; by \lq\lq
critical region\rq\rq\/
we understand the region in which $\rho\simeq 1$.

As was discussed in Ref. [\onlinecite{amp}], a maximum in
$\rho(T)$ results from a competition between metallic-like
$\rd(T)$ and localization effects, the latter being controlled
both by $\rd(T)$ and $\tau_{\varphi}$. At higher $T$, when $\rd$
is sufficiently large, localization can already be strong enough
to ensure the negative sign of $d\rho/dT$ despite
$\rd(T)$ decreasing with temperature.
However, as $T$ goes down, $\rd$ decreases, localization  weakens and
$d\rho/dT$ changes its sign from negative to positive, thus a
maximum in $\rho(T)$ occurs.
This can happen provided that certain conditions are satisfied.
In particular, one has to require that
\beq
\rd(T_0)>\rho_c,
\eeq
where $T_0$ is the temperature at which  $\tau_{\varphi}$
is comparable to the transport mean free time  (the flow of
Eq.~(\ref{rgrho}) starts at $T=T_0$) and $\rho_c$ is some critical
resistivity determined
by a particular form of $\rd(T)$.
For example, if $\rd\propto T^q$, critical resistivity $\rho_c$ is a
certain function of $p/q$ (see Eq.~(9) of Ref.~[\onlinecite{amp}]). In
this case, there is another condition for the maximum in $\rho (T)$ to
occur, namely, $p>2q$.

As the resistivity $\rd(T_0)$ depends on the electron density
$n$, it can be tuned by the gate voltage.
This tuning can drive the system between the domains of parameters
corresponding either to a maximum or no maximum in $\rho(T)$.
Let us start at $\rd(T_0)>\rho_c$, so that the maximum in $\rho$ is at
$T=\tmax$.
As $\rd$ approaches $\rho_c$ from above, $\tmax$  remains finite, while
the maximum flattens out and disappears
as soon as
$\rd(T_0)$ approaches $\rho_c$.
At this moment, both $d\rho/dT$ and $d^2\rho/dT^2$ vanish, and the
variation of $\rho$ around $\tmax$ takes place solely due to higher
($n\geq 3$) terms in the Taylor expansion of $\rho$.

How large is the temperature interval, in which this variation does not
exceed some given value?
Denoting,
\bea
t&\equiv&\ln T,\nonumber\\
\tm &\equiv & \ln\tmax,\nonumber\\
\bd & \equiv & \beta[\rd(\tm)],\nonumber\\
y(t)&\equiv & \ln\rd(t), \nonumber\\
\bd' &\equiv & d\beta(\rho)/d\ln\rho\Big|_{\rho=\rd}, \nonumber\\
\bd'' &\equiv & d^2\beta(\rho)/d\ln\rho^2\Big|_{\rho=\rd},\nonumber
\eea
we obtain from Eq.~(\ref{rgrho})
\begin{mathletters}
\bea
\frac{d\rho}{dT}\Big|_{T=\tmax}=0&\Rightarrow & {\dot y}(\tm)=\frac{p}{2}
\frac{\bd}{\bd-1}\label{rdot}\\
\frac{d^2\rho}{dT^2}\Big|_{T=\tmax}=0&\Rightarrow &{\ddot y}(\tm)=
-\frac{\bd'}{\bd(\bd-1)}\left[{\dot y}(\tm)\right]^2.
\label{rddot}\eea
\end{mathletters}
The third (logarithmic) derivative of $\rho$ is then found from
Eq.~(\ref{rgrho}) to be
\beq
\frac{d^3\ln\rho}{dt^3}\Big|_{t=\tm}=(1/A)\left[S-
\frac{\bd-1}{\bd}\frac{d^3 y}{dt^3}\Big|_{t=\tm}\right],
\eeq
where
\beq
A\equiv 1/\beta(\rho)-\gamma
\eeq
and
\begin{mathletters}
\bea
S&\equiv & \frac{2\bd'^2-\bd\bd''}{\bd}^3 \left[{\dot y}(\tm)\right]^3-
\frac{3\bd'}{\bd^2}{\ddot y}(\tm){\dot y}(\tm)\label{eqSa}\\
&=&\frac{p^3}{8(\bd-1)^4}\left[\left(2\bd^{'2}-\bd\bd^{''}\right)(\bd-1)+
3\bd^{'}\right]\label{eqSb}.
\eea
\end{mathletters}
(In going from (\ref{eqSa}) to (\ref{eqSb}), we used
(\ref{rdot},\ref{rddot})).

\begin{figure}
\begin{center}
\resizebox{3.2in}{4.3in}{\includegraphics{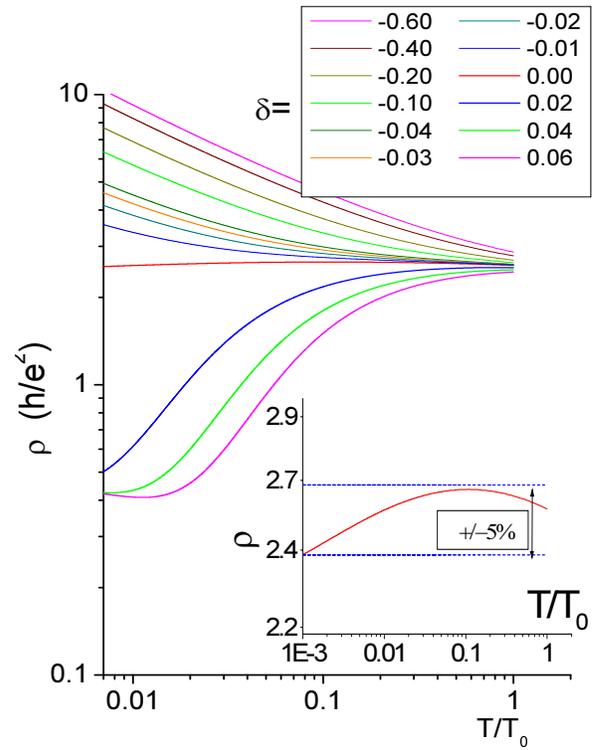}}
\begin{minipage}{3in}
\vspace{0.1in}
\caption{Solution of Eq.~(\ref{rgrho}) for different values of $\delta$
(distance from the \lq\lq critical point\rq\rq\/). $p=1, q=1/4, \gamma=1,
\rho_1=0.28, {\bar \rho}=2.3$.
Inset: the central (almost flat) curve ($\delta=0$) over a wider
temperature range.}
\label{fig6}
\end{minipage}
\end{center}
\end{figure}

For a sufficiently narrow interval of $t$ around $\tm$, it suffices to
keep only the lowest (cubic) term in the Taylor expansion of
$\ln \rho (T)$ as a function of $\ln (T/\tmax )$:
\bea
\ln \left(\frac{\rho(t)}{\rho(\tm)}\right)
=&\left(1/6A\right)\left[S-\frac{\bd-1}{\bd}\frac{d^3\ln\rd}
{dt^3}|_{t=\tm}\right]\times & \nonumber \\
&\times\left(t-\tm\right)^3.
\label{est}
\eea
As $y$ is supposed to be a smooth function of $t$, one can
estimate $d^n y/dt^n$ as ${\bar y}/{\bar t}^n$, where ${\bar y}$
is the typical value of $y$ in the interval ${\bar t}$.
In the \lq\lq critical region\rq\rq\/, $\rho\simeq 1$,
which means that ${\bar y}\simeq 1$ as well, and hence
$\bd\simeq\bd'\simeq\bd''\simeq 1$.
Thus the two terms in the square brackets in (\ref{est}) are of the same
order and (\ref{est}) reduces to
\beq
\ln\rho(t)-\ln\rho(\tm)\simeq \left(S/6A\right)(t-\tm)^3.
\label{est1}
\eeq
Let us now estimate the number $D$ of decades in temperature
\beq
T_{{\rm
high}}/T_{{\rm low}}=10^D,
\eeq
for which
\beq
|\ln\rho(t)-\ln\rho(\tm)|\leq 10\%.
\eeq
(KK found that their $\rho$ is $T$-independent with this accuracy.)
Using an interpolation formula
$\beta(\rho)=-\ln\left(1+2\rho/\pi\right)$
[\onlinecite{isa}] and choosing $\rd(\tmax)=\pi/2$,
we have
\beq
D\lesssim 2.3|A|^{1/3}/p.
\label{decade}
\eeq
For $p=1$ and $|A|=1$, we get $D\lesssim 2.3$, i.e., the resistivity stays
within a 10\%-margin for more than {\it two decades} in $T$.
(Note that a numerical coefficient of the order of unity,
which we neglected in deriving (\ref{est1}), would enter (\ref{decade})
only under the cubic root, and is thus unimportant.)
Fig.~\ref{fig6} demonstrates $\rho(T)$ calculated from
Eq.~(\ref{rgrho}) for $\rd$ taken from the model of
Ref.~[\onlinecite{am}]:
\bea
\noindent
\rho _{{\rm d}}&=&\rho _{1}+{\bar\rho}\left( T/T_{0}\right)
^{q}\nonumber \\
&&\times\left\{
\begin{array}{ll}
\left( 1+|\delta |T_0/cT\right) ^{q},\;{\rm for}\;\delta \leq 0,\,{\rm
``insulator''}; &  \\
e^{-\delta T_0/T},\;{\rm for}\;\delta >0,\,{\rm ``metal''}, &
\end{array}
\right.
\label{rdtrap}
\eea
where $\delta=\delta_0(n-n_c)/n_c$ characterizes the distance
from the critical point and $c\simeq 1$. The dimensionless
coefficient $\delta_0$
depends on details of the model (cf. Eq.(11) of
Ref.~[\onlinecite{amp}]).
For the purposes of the present paper, we view Eq.~(\ref{rdtrap}) as
simply a
phenomenological form for $\rd(T)$,
regardless of the model [\onlinecite{am}] it was  originally derived from.
This form is consistent with the experiment, if $\delta_0\simeq 1$.

The value of $\delta=0$ corresponds to the central line in Fig.~\ref{fig6},
 which on this scale is
essentially constant for more than two decades.
Zooming in (cf. the inset in Fig.~\ref{fig6}),
one sees that in fact $\rho(\delta=0,T)$ stays within a 10\%-margin over
exactly {\it three decades} in $T$.
\vspace{-0.1in}
KK claim that the observed $\rho(T)$ stays within this margin
over only 1.5 decades, which is smaller than even the conservative
estimate (\ref{decade}).
(As it was discussed in Sec.~\ref{sec:heating}, a more realistic estimate
of KK's $T$-interval is one decade).
A decrease in $\delta$ by 1\% ($\delta=-0.01$) leads
to a weakly insulating behavior: $\rho(T)$ increases
by 30\% as $T$ decreases by 1.5 decades. An increase
in $\delta$ by 4\% ($\delta=0.04$) leads to a well-pronounced
metallic behavior: $\rho(T)$ drops by about a factor
of two over 1.5 decades in $T$. These features are
in quantitative agreement with the experiment [\onlinecite{kk}].
(For the sake of simplicity, we assumed
that parameters $\rho_1$ and ${\bar \rho}$ do not depend
on the density, i.e., on $\delta$.
Taking these dependences into account should lead to even better agreement
with the experiment.)

We emphasize that {\it no fine tuning} between the $\beta$-function and
$\rd$ was used.
If it is the competition between the decrease in Drude
resistivity with temperature and the localization effects that leads to
the maximum in the observable resistivity,
then
the $\rho (T)$-dependence flattens out, as the density approaches the
threshold for a maximum.
It turns out that if the exponent $p$ is not too big (which is usually
the case), this flattening suppresses the variation of $\rho (T)$  below
the experimentally observable level over several decades in temperature.
Even an oversimplified model [\onlinecite{amp}]
which does not take into account, e.g., electron-electron interactions (except
for as a possible phase-breaking mechanism),
can
easily produce almost constant resistivity in a temperature interval,
which is {\it two orders of magnitude} wider than the experimental one.
Therefore, we do not find it  too much of
a \lq\lq remarkable coincidence \rq\rq\/
that at some density the system demonstrates a fairly constant
$\rho$.

We now turn to the data presented in Ref.~[\onlinecite{kk}].
The authors emphasize strongly that at $ n=n_{3}=7.25 \times
10^{11}$\,cm$^{-2}$ the resistivity
is almost temperature independent
in the temperature range 35\,mK - 1\,K.
They also
add that $\rho(T)$ decreases with $T$ for $T\gtrsim 1.8$\,K.
It would be interesting to know what happens at
intermediate temperatures: $1\,{\rm K}<T<1.8$\,{\rm K}.
This point being unclear,
KK
fill in the missing temperature interval
by combining their result with the one obtained on another
system.
Indeed, they write:
\\
\\
\noindent
\lq\lq In combination with the results of Ref.~[10]
(Ref.~[\onlinecite{academy}]
of this paper--AMP) where the temperature-independent curve (with
essentially
the same value of resistivity) was observed
in the temperature range
250 mk-1.8\,K in another 2D system in silicon, we further allege that
there
is no observable $T$-dependence at $n_s=n_c$ in the temperature range
35\,mk-1.8 K...\rq\rq\/.
(Ref.~[\onlinecite{kk}], p.~3, second paragraph.)
\\
\\
Parenthetically, we note that the resistivities of combined curves
differ by 35\%.
As far as we understand,
temperature intervals of experiments on different systems
are not additive parameters,
and thus the procedure described above is not justified.

Summarizing this part of our discussion, we take the liberty to describe
the experimental situation in the following way:

In a sample from the intermediate peak mobility range (where temperature
dependence of $\rho$ is known to be quite weak at high $T$), one can find
a density such that $\rho$  does not change for more than 10\% within
about one order of magnitude in $T$.  This
statement is a result of measurements on a single sample in the interval
100\,mK$<T<1$\,K  (when a realistic estimation of the electron heating is
taken into account).
(It will be two samples if results of
Ref.~[\onlinecite{academy}] for
250mK$<T<$1.8K are taken into consideration.)

In our opinion, there are neither theoretical
 nor experimental reasons to believe that the
density corresponding to the weakest $\rho(T)$-dependence is indeed the
critical one. i.e., that $\rho (T\to 0) \to \infty$ and $\rho (T\to 0)$ is
finite for lower and higher densities, respectively.
The lack of a pronounced temperature dependence of $\rho$
in a limited range of $T$
does not signal any remarkable
phenomenon and is well-described by a simple theory
of Anderson localization in a temperature-dependent disorder
(Ref.~[\onlinecite{amp}]).

\section{On the apparent linear temperature dependence of $\rho(T)$}
\begin{figure}
\begin{center}
\resizebox{3.2in}{3.in}{\includegraphics{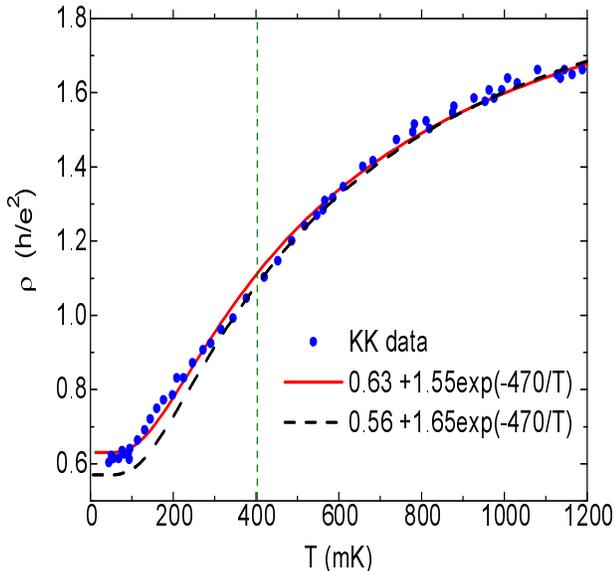}}
\begin{minipage}{3in}
\caption{Temperature
dependence reproduced from Figs.~1, 3 of  Ref.~
[15].
 Region to the left of the vertical
dashed line is the truncated $T$-range in Fig.~3a of
Ref.~[15], in which
$\rho (T)$ is claimed to be linear.
Bold dashed and continuous curves: fits of the data with
Eq.~(\ref{empiric}).}
\label{fig7}
\end{minipage}
\end{center}
\end{figure}

We now turn to point (2) of
Ref.
[\onlinecite{kk})]  regarding
the functional form of the observed $\rho(T)$-dependence.
 According to KK, this form is a)~linear and b)~qualitatively
distinct from those observed by other authors.
Fig.~\ref{fig7} reproduces Fig.~3b
of Ref.~[\onlinecite{kk}], where the
\lq\lq almost linear dependence\rq\rq\/
is demonstrated.
The data in Fig.~\ref{fig7} is data {\it 5}
of Fig.~\ref{fig1}, displayed  in truncated ($T<400$\,K)
and full ($T<1200$\,mK) intervals.
The truncated interval, in which $\rho(T)$ is supposed
to be \lq\lq almost linear\rq\rq\/,  corresponds
to Fig.~3b of Ref.~[\onlinecite{kk}].

We begin with an attempt to fit untruncated data {\it 5} of
Figs.~\ref{fig1}, \ref{fig7} in the
whole interval $T<1.2$\,K
 by an
 empirical expression [\onlinecite{pudalov97}]
\beq
\rho(T) = \rho_0 + \rho_1 \exp\left(- \frac{T_0}{T} \right),
\label{empiric}
\eeq
which is a simplified version of
Eq.~(1) with $\alpha =0$. This expression is known to work reasonably
well
for the metallic phase not only in Si MOSFETs
[\onlinecite{noscaling,pudalov97}] but also in other 2D systems
exhibiting the MIT-like behavior [~\onlinecite{shahar}].

As is seen from Fig.~\ref{fig7}, this attempt is quite successful.
We achieved more than just acceptable fit in a much broader temperature
interval than the one in which $\rho(T)$ looks
\lq\lq almost linear\rq\rq\/.
Of course, it is not a much of an achievement to fit a smooth curve by a
function with three free parameters (to demonstrate a flexibility in
choosing the fitting parameters,
Fig.~\ref{fig7} presents two  sets of parameters that can be used).  On
the other hand, we see {\it no reason} to
argue that the $\rho(T)$-dependence of Ref. [~\onlinecite{kk}] at
$n=n_{4}$ and $n=n_{5}$ (see Fig.~1) is qualitatively different from
those observed in other experiments.

Moreover, such \lq\lq almost linear\rq\rq\/ $\rho(T)$-dependences are
{\em typical for low-mobility samples in the \lq\lq
metallic\rq\rq\/ regime}.
For instance, compare Fig.~\ref{fig7} with Fig.~\ref{fig5}b, in
which $\rho(T)$ for a  low-mobility sample is plotted over a much wider
temperature range (90\,mK - 4\,K), and for $n$ about 50\% higher than what
can be called a critical concentration.
The similarity of these two plots is quite clear.

We find it quite unlikely that the quasi-linear $\rho(T)$-dependences of
Figs.~\ref{fig1} and \ref{fig7}
(as well as Fig.~5) manifest an exotic non-fermi-liquid state.
Taking into account the whole set of features of the metallic state in
Si-MOSFETs (which we listed in Ref.~[\onlinecite{amp}]),
we conclude that these
dependences must have a classical explanation (see the next paragraph for
our
definition of this term).
For example, one cannot exclude that such behavior is
caused by the temperature dependence
of the screening length [\onlinecite{dolgo,das,klapwijk,das99}]
or by the temperature dependence of the
charge traps' population in the oxide [\onlinecite{am}].

We conclude this discussion by remarking that there seems
to be some confusion in recent literature on MIT
in 2D (see, e.g., Ref.~[\onlinecite{sachdev}]) regarding the distinction
between \lq\lq quantum\rq\rq\/ and \lq\lq classical\rq\rq\/ behaviors.
We believe that this confusion is mainly semantic. We call a regime
\lq\lq classical\rq\rq\/ as long as
electron transport can be  described in the framework
of the Boltzmann theory. This does not necessarily  mean
that electrons obey Boltzmann (non-degenerate) statistics
-in fact, the Fermi-liquid theory was originally
 formulated in terms of the
Boltzmann equation [\onlinecite{landau}].
It also does not exclude that a scattering cross-section
is described in terms of quantum mechanics. (A more
common name for this regime is \lq\lq semi-classical\rq\rq\/
[\onlinecite{ashcroft}],
although the semi-classical approximation may not necessarily
be employed for the calculation of a scattering cross-section.)
However, quantum-mechanical effects arising due to interference
of electron waves, e.g., weak localization, cannot be described
by the Boltzmann equation. We call the regime
\lq\lq quantum\rq\rq\/,
if these effects are important.
Although we agree with the authors of Ref. [~\onlinecite{sachdev}]
in that \lq\lq the formation of a Fermi surface\rq\rq\/
is an \lq\lq intrinsic quantum-mechanical effect\rq\rq\/,
this does not imply that the metallic-like temperature dependence
of the resistivity, whose onset is correlated with the
Fermi-surface formation, is of the quantum nature,  in a sense of
the definition given above.

\section{At what temperatures weak localization may be observed}
There is one more point to discuss regarding the metallic behavior of
Si MOSFETs.  In addition to the almost linear dependence with
$d\rho/dT>0$, KK emphasize that no signs of weak localization
(WL) was observed, although they went to temperatures as low as
$T\sim 0.008T_F$, where $T_F$ is the Fermi temperature.
Giving the authors the benefit of the doubt, we assume
the most favorable scenario in which electrons are not overheated.
Nevertheless, under this assumption the  apparent
absence of WL is not  as
surprising as KK present it.
In fact, we believe that even the {\it bath} temperature was too high to
observe the localization upturn in the $T$-dependence of the resistivity.
Indeed, experimental studies of high-mobility Si MOSFETs in the wide
range of electron densities [~\onlinecite{JETPL97,weakloc}] show that
WL effects are well-pronounced in the (perpendicular)
{\it magnetic field} dependence of the resistivity.
The observed magnetoresistance is in a quantitative agreement with the
conventional theory [\onlinecite{leeandrama,aa}], even for densities
rather close to the critical one.
On the other hand, even for $n\gg n_c$
observation of the WL
effect
in the zero-field
$\rho(T)$-dependence [\onlinecite{JETPLlnT,gmax}]
(as well as the effects of electron-electron interactions)
requires a precision of the order of
$\delta \rho(T)/\rho(T) \ll (1/\pi)(e^2/h)\rho \sim 0.1\%$.

It turns out that the observed dependence
in the range of \lq\lq high temperatures\rq\rq\/, $T=(0.1 - 0.5)E_F$,
is too strong to be explained by
WL and quantum interaction effects, and can be attributed
to the $T$-dependence of the Drude resistivity.
However, the slope $|d\rho/dT|$ decreases
with
$T$.
For relatively large densities, one can reach the temperature region,
$T<T_Q$, in which the slope becomes comparable to that of the WL
correction, $|d\rho/dT|_{WL}\sim (e^2/\pi)\rho^2T$.
In this region, quantum interference should contribute significantly to
the $\rho(T)$-dependence.
Authors of Ref.~[\onlinecite{gmax,JETPLlnT}]
found an
empirical relation: $T_Q\lesssim
0.007 E_F$ (see Fig.(\ref{fig2})).
As $E_F$ is proportional to $n$, it becomes
progressively more challenging to cool electrons below $T_Q$ as $n_c$ is
approached.
$E_{F} \simeq 5.5$\,K for the highest electron density reported by
KK and thus $T_Q=40$\, mK.
Therefore, in order just to reach the high temperature edge  of the
quantum transport region, where WL
effects are seen, electrons should  be cooled below 40\,mK.
The localization upturn may be expected to occur only at temperatures
substantially smaller than even this one.
In other words, the lowest temperature reported by KK (35\,mK) is still
too high for localization
effects to be observed.

Why WL manifests itself so differently in the magnetoresistance and in
the $\rho(T)$?
This can be explained naturally by assuming that the Drude resistivity
has a pronounced $T$-dependence, which masks quantum corrections.
On the other hand,  $\rd$ is not expected to
vary substantially
in weak magnetic fields, thus WL can be seen in magnetoresistance.

\section{Conclusions}
\label{sec:concl}
Some concluding remarks:
\begin{enumerate}
\item[i)]
the apparent metal-insulator transition in 2D
is a very interesting and unexpected phenomenon.
Although we have limited our discussion to Si MOSFETs,
which exhibit the
strongest, among other 2D systems,
\lq\lq anomalous\rq\rq\/ metallic behavior, there is
a whole variety of interesting effects not only
in transport but also thermodynamic properties,
e.g., compressibility [\onlinecite{jiang,yacoby}]),
observed in
various
2D heterostructures.
Whether all these observations have a universal
explanation or there is a number of different,
system-specific mechanisms at work, remains to be seen.
\item[ii)] we do not believe that the experimental
results accumulated up-to-date provide a convincing
evidence for this phenomenon being a true quantum
phase transition between two distinct states of matter.
More precisely, there is no evidence that
the \lq\lq metallic phase\rq\rq\/ (as defined by the sign
of $d\rho/dT$) is a new state of matter.
On the contrary, a whole set of features--see
Refs.[\onlinecite{amp,meir1}]--is consistent
with a conventional behavior of a disordered Fermi liquid.
This statement does not negate i).
There is still no consensus on the origin of the effect, and this
problem requires the most serious and intensive investigations.
\item[iii)] For  a successful resolution of
the problem, it is probably not sufficient to concentrate on the
temperature dependence of the resistivity in the vicinity of the
transition. Such features  as a weak temperature dependence of the
resistivity at a certain electron density in a limited (though large)
temperature interval allow for different interpretations and,
unfortunately, can not provide an unambiguous information about the
zero-$T$ state of the system.
In particular, it is impossible to determine even the critical density
without making rather arbitrary assumptions.
\item[iv)] There is a serious experimental difficulty that up to now
prevented a substantial increase of the temperature interval, in which
the resistivity can reliably be
measured in Si-MOSFETs.
It turns out that in existing samples phonons can hardly
cool the electron gas below a fraction of a Kelvin. On the
other hand,
the systems are quite noisy, and thus the applied voltage cannot be
reduced much below the currently used level.  As a result, even in the
absence of non-equilibrium noise, one cannot neglect electron heating
at $T \simeq 100$\,mK.
\end{enumerate}
\acknowledgements The work at Princeton University was supported
by ARO MURI DAAG55-98-1-0270. D.\ L.\ M. acknowledges the
financial support from NSF DMR-970338 and from the Research
Corporation Innovation Award (RI0082). V.\ M.\ P. acknowledges the
support from RFBR, INTAS, NWO, NATO (PST.CLG.976208) and  Programs
\lq\lq Physics of solid-state nanostructures\rq\rq\/,
\lq\lq Statistical Physics\rq\rq\/, and \lq\lq Integration\rq\rq\/.
We would like to thank the Center of
Higher Studies (Oslo, Norway) and
Institut f\"{u}r
Halbleiterphysik Johannes Kepler Universit\"{a}t (Linz, Austria),
where parts of this work were done. We are grateful to  M.\ Reizer
for numerous illuminating discussions and to
A.\ F.\ Hebard, R.\
Fletcher, C.\ Marcus, G.\ W.\ Martin, and A.\ V.\ Varlamov for  critical
reading of the manuscript and valuable comments. We also appreciate
the
assistance of  G.\ W.\ Martin in the manuscript preparation.

\end{multicols}

\begin{references}
\bibitem{comment}
We wish to emphasize that what we call ``Drude resistivity''\/
is not necessarily something well-known and described in
textbooks. It is quite possible that the mechanism(s) of the
unusual Drude resistivity in Si MOSFETs and other 2D systems is
(are) based on new physics. The only one feature of the Drude
resistivity which we retain in our argument is that $\rd$ has
nothing to do with quantum interference.
\end{references}
\end{document}